\documentclass[runningheads]{cl2emult}

\usepackage{makeidx}  
\usepackage{graphicx} 
\usepackage{epsfig}
\usepackage{subeqnar} 
\usepackage{multicol} 
\usepackage{cropmark} 
\usepackage{lnp}      
\usepackage{amsmath}
\usepackage{latexsym}
\usepackage{citesort}
\makeindex            

\begin{document}
\title*{The equation of state for almost elastic, smooth, 
        polydisperse granular gases for arbitrary density}
\toctitle{The equation of state of polydisperse granular gases}
\titlerunning{Polydisperse granular gases}
\author{Stefan Luding
\and Oliver Strau\ss{}}
\authorrunning{Stefan Luding and Oliver Strau\ss{}}
\institute{
   Institute for Computer Applications 1, \\
   Pfaffenwaldring 27, 70569 Stuttgart, Germany.
   e-mail: lui@ica1.uni-stuttgart.de }

\maketitle

\begin{abstract} 
Simulation results of dense granular gases with particles of different 
size are compared with theoretical predictions concerning the 
pair-correlation functions, the collison rate, the energy dissipation,
and the mixture pressure. The effective particle-particle correlation 
function, which enters the equation of
state in the same way as the correlation function of monodisperse
granular gases, depends only on the total volume fraction and on
the dimensionless width ${\cal A}$ of the size-distribution function.
The {\em global equation of state} is proposed, which unifies both the 
dilute and the dense regime.

The knowledge about a global equation of state is applied to steady-state
situations of granular gases in the gravitational field, where averages 
over many snapshots are possible. The numerical results on the density 
profile agree perfectly with the predictions based on the global equation 
of state, for monodisperse situations. In the bi- or polydisperse cases,
segregation occurs with the heavy particles at the bottom.
\end{abstract}

\section{Introduction}

The hard-sphere (HS) gas is a traditional, simple, tractable model
for various phenomena and systems like e.g. disorder-order 
transitions, the glass transition, or simple gases and liquids 
\cite{chapman60.Luding,ziman79,hansen86}. A theory that describes the 
behavior of rigid particles is the kinetic theory 
\cite{landau86,chapman60.Luding}, where particles are assumed to be rigid 
and collisions take place in zero time (they are instantaneous),
exactly like in the hard-sphere model. 
When dissipation is added to the HS model, one has the most simple version 
of a granular gas, i.e.~the inelastic hard sphere (IHS) model.
Granular media represent the more general class of dissipative, 
non-equilibrium, multi-particle systems \cite{herrmann98}.  
Attempts to describe granular media by means of kinetic theory
are usually restricted to certain limits like constant or small density
\cite{haff83.Luding} or weak dissipation \cite{jenkins85b.Luding,sunthar99}. Also in the
case of granular media, one has to apply higher order corrections
to successfully describe the system under more general conditions
\cite{noije98c,sela98}.  Already classical numerical studies 
showed that the equation of state can be expressed as some power series
of the density in the low density regime 
\cite{alder59,ziman79,hansen86}, whereas, in the high density
case, the free volume theory leads to useful results \cite{buehler51}.
In the general situation, the granular system consists of particles
with different sizes, a situation which is rarely addressed theoretically
\cite{jenkins87,arnarson98,arnarson99,willits99}. However,
the treatment of bi- and polydisperse mixtures is easily
performed by means of numerical simulations 
\cite{dickinson77,luding99b,mcnamara99}.

In this study, theories and simulations for situations with 
particles of equal and different sizes are compared. In section 
\ref{sec:model} the model system is introduced and in
\ref{sec:simtheo} we review theoretical results and compare them
with numerical results concerning correlations, collision rates,
energy dissipation and pressure. Based on the numerical data,
a global equation of state is proposed. This global equation of
state is valid for arbitrary densities and mixtures with particles
of different size, and it is used to explain
the density profile in a dense system in the gravitational field
in section\ \ref{sec:pgrad}. The results are summarized and discussed 
in section\ \ref{sec:concl}.

\section{Model system}
\label{sec:model}

For the numerical modeling of the system, periodic, two-dimensional (2D) 
systems of volume $V=L_x L_y$ are used, with horizontal and vertical size 
$L_x$ and $L_y$, respectively.  $N$ particles are located at positions
${\vec r}_i$ with velocities $\vec{v}_i$ and masses $m_i$.
From any simulation, one can extract the kinetic energy
$E = \frac{1}{2} \sum_{i=1}^N m_i \vec{v}_i^2$, dependent on time
via the particle velocity $\vec{v}_i$. In 2D, the ``granular temperature'' 
is defined as $T=E/N$.

\subsection{Polydispersity}

The particles in the system have the radii $a_i$ 
randomly drawn from size distribution functions $w(a)$ as summarized in 
table\ \ref{tab:sdf} where the step-function $\theta[x]=1$ for $x \ge 1$ 
and $\theta[x]=0$ for $x < 1$ is implied.
\begin{table}
\begin{center}
\begin{tabular}{|r|l|l|}
\hline
  (i) $~~$ & monodisperse   & $~~$ $w(a)=\delta(a-a_0)$ ~ \\
\hline
 (ii) $~~$ & bidisperse     & $~~$ $w(a)=n_1 \delta(a-a_1) + n_2 \delta(a-a_2)$ ~ \\
\hline
(iii) $~~$ & polydisperse   & $~~$ $w(a)=\frac{1}{2 w_0 a_0} \theta[a-(1-w_0) a_0] \, \theta[(1+w_0) a_0 -a] ~~$ \\
\hline
\end{tabular}
\end{center}
\caption{Size distribution functions used in this study.}
\label{tab:sdf}
\end{table}

~\vspace{-.8cm}\\
The parameter $a_0$ is the mean particle radius $\langle a \rangle$ 
in cases (i) and (iii). In the bidisperse situation (ii), one has 
$a_0 = \langle a \rangle = n_1 a_1 + n_2 a_2 = (n_1+(1-n_1)/R)a_1$, 
with the fraction  $n_1 = N_1/(N_1+N_2)$ of particles with size $a_1$ 
in a system with $N=N_1+N_2$ particles in total and $N_2$ particles
with radius $a_2$. Thus, besides $n_1$, only the size ratio $R=a_1/a_2$ is
needed to classify a bidisperse size distribution. The total 
volume fraction $\nu = \nu_1 + \nu_2$ is the last relevant system parameter,
since the partial volume fractions $\nu_{1,2} = N_{1,2} \pi a_{1,2}^2/V
= n_{1,2} \nu a_{1,2}^2 / \langle a^2 \rangle$ can be expressed in terms 
of $n_1$ and $R$: Using the dimensionless moments 
\begin{equation}
A_k = n_1 + (1-n_1) R^{-k} = \frac{\langle a^k \rangle }{ a_1^k }~,
\end{equation}
one has $\nu_1 = n_1 \nu / A_2$ and $\nu_2 = (1-n_1) \nu / (R^2 A_2)$.
Since needed later on, the expectation values for the moments of $a$
and their combination, the dimensionless width-correction ${\cal A} =
\langle a \rangle^2/\langle a^2 \rangle$, are
summarized in table\ \ref{tab:moments} in terms of $a_1$, $n_1$, and $R$ 
for the bidisperse situations and in terms of $a_0$ and $w_0$ in the 
polydisperse cases. Different values of $\nu$ are realized by shrinking 
or growing either the system or the particles.
\begin{table}
\begin{center}
\begin{tabular}{|r|l|l|l|l|}
\hline
  &  &  $~~$ $\langle a \rangle$ &  $~~$ $\langle a^2 \rangle$  &
        $~~$ $\langle a \rangle^2/\langle a^2 \rangle$ \\
\hline
\hline
  (i) $~~$ & monodisperse   & $~~$ $a_0$  
      & $~~$ $a_0^2$ 
      & $~~$ $1$ \\
\hline
 (ii) $~~$ & bidisperse     & $~~$ $A_1 a_1$
      & $~~$ $A_2 a_1^2$ 
      & $~~$ ${A_1^2}/{A_2}$ \\
\hline
(iii) $~~$ & polydisperse   & $~~$ $a_0$  
      & $~~$ $\left ( 1+{w_0^2}/{3} \right ) a_0^2$ 
      & $~~$ 3 / $\left ( {3+w_0^2} \right )$ \\
\hline
\end{tabular}
\end{center}
\caption{Moments $\langle a \rangle$, $\langle a^2 \rangle$ and 
${\cal A}=\langle a \rangle^2/\langle a^2 \rangle$ of the size 
distribution functions.}
\label{tab:moments}
\end{table}

~\vspace{-1.5cm}\\
\subsection{Particle Interactions}

The particles are assumed to be perfectly rigid and to follow an 
undisturbed motion until a collision occurs as described below.
Due to the rigidity, collisions occur instantaneously, so that an event 
driven simulation method  \cite{lubachevsky91,luding98f} can be used.
Note that no multi-particle contacts can occur in this model. For
a review on possible, more physical extensions of this model see
Ref.\ \cite{luding98f}.  

A change in velocity -- and
thus a change in energy -- can occur only at a collision.
The standard interaction model for instantaneous collisions of
particles with radii $a_i$, mass $m_i=(4/3)\pi \rho a_i^3$,
and material density $\rho$ is used in the following.
(Using the mass of a sphere is an arbitrary choice, however, using
disks would not influence most of the results discussed below.)
This model was introduced and also discussed for the more
general case of rough particle surfaces in Refs.\
\cite{jenkins85b.Luding,walton86b,lun91.Luding,Goldshtein95.Luding,luding98d}.
The post-collisional velocities $\vec{v}'$ of two collision partners
in their center of mass reference frame are given, in terms of
the pre-collisional velocities $\vec{v}$, by
\begin{equation}
\vec{v}_{1,2}' = \vec{v}_{1,2} \mp \frac{(1+r)~m_{12}}{m_{1,2}} \vec{v}_n ~,
\label{eq:collrule}
\end{equation}
with $\vec{v}_n \equiv \left [ (\vec{v}_1 - \vec{v}_2) \cdot \hat{\vec{n}} \right ] \hat{\vec{n}}$,
the normal component of $\vec{v}_1-\vec{v}_2$ parallel to $\hat{\vec{n}}$,
the unit vector pointing along the line connecting the
centers of the colliding particles, and the reduced mass
$m_{12} = m_1 m_2 / (m_1+m_2)$.
If two particles collide, their velocities are changed according
to Eq.\ (\ref{eq:collrule}) and any $r(v_n)$, dependent on $v_n = |\vec{v}_n|$,
can be used.  For a pair of particles, the change of the translational
energy at a collision is $\Delta E = - m_{12} (1-r^2) v_n^2 / 2$.

\section{Simulation and theory}
\label{sec:simtheo}


In the following, we compare simulations with different polydispersity,
i.e.~different size distribution functions $w(a)$, as summarized in 
table\ \ref{tab:simpar}.
\begin{table}
\begin{center}
\begin{tabular}{|l|l|l|l|r|}
\hline 
  &              & $w(a)$ parameters    & particles & ${\cal A}$ \\ 
\hline 
\hline 
A &monodisperse  & $w_0=0$,             & $N=1628$  & 1 \\ 
\hline
B &monodisperse  & $n_1=1$, $R=1$ & 
                                          $N=576$   & 1 \\ 
\hline
C &bidisperse    & $n_1=0.517$, $R=3/4$ & 
                                          $N=576$   & ~0.9798 \\ 
\hline
D1 &bidisperse    & $n_1=0.781$, $R=1/2$ & 
                                          $N=576$   & ~0.8968 \\ 
D2 &bidisperse    & $n_1=0.799$, $R=1/2$ & 
                                          $N=6561$  & ~0.8998 \\ 
\hline 
E &polydisperse   & $w_0=0.25$           & $N=1425$  & ~0.9796 \\ 
\hline 
F1 &polydisperse  & $w_0=0.5$            & $N=1425$  & ~0.9231 \\ 
F2 &polydisperse  & $w_0=0.5$            & $N=1521$  & ~0.9231 \\ 
\hline 
\end{tabular}
\end{center}
\caption{Simulation parameters for the simulations discussed below. Note that
sets C and E have different $w(a)$ but almost identical values of ${\cal A}$.}
\label{tab:simpar}
\end{table}

~\vspace{-1.4cm}\\
\subsection{Particle correlations}

\begin{figure}[ht]
\begin{center}
\hspace{-.4cm}
\psfig{file=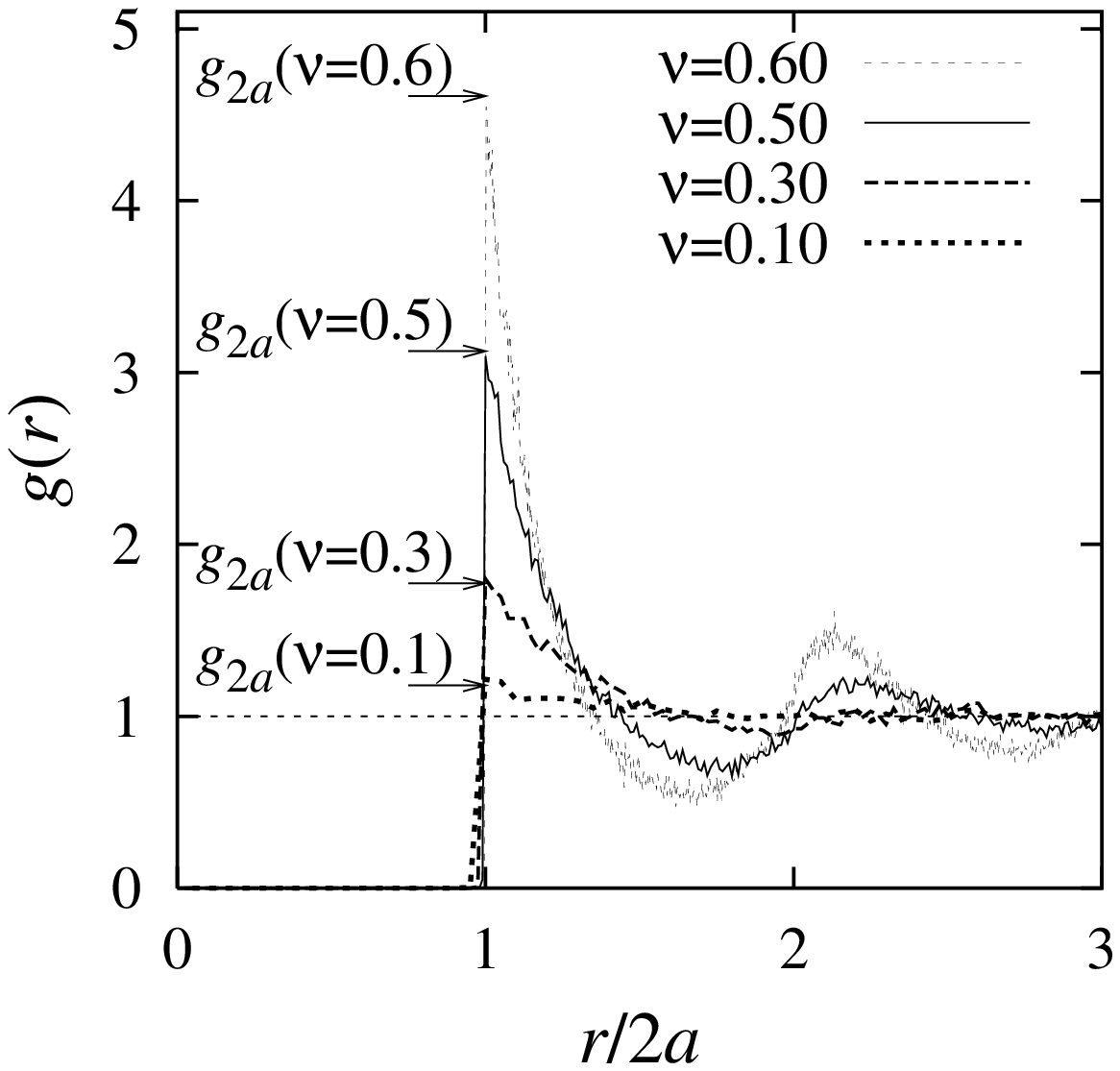,width=5.6cm,angle=-90} \hspace{-.8cm}
\psfig{file=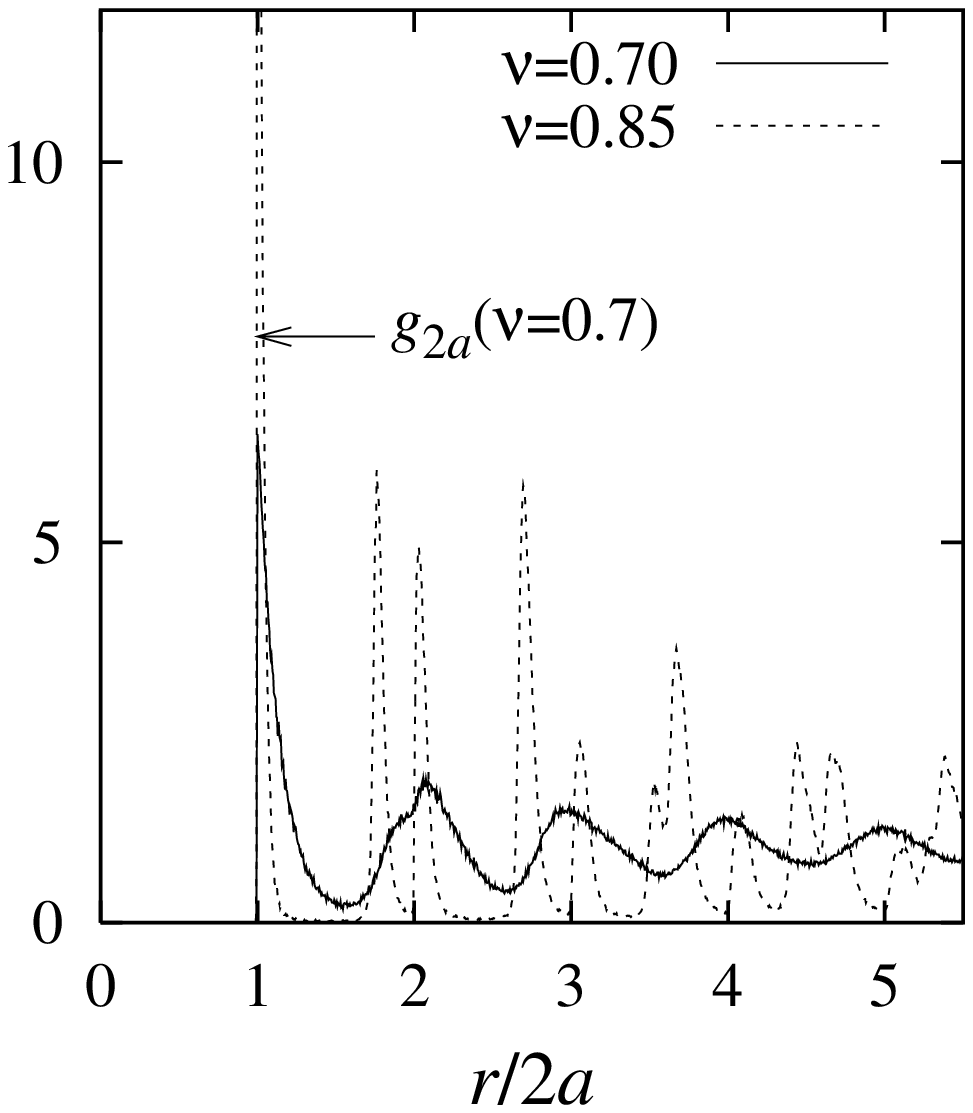,width=5.6cm,angle=-90}\\ 
~\vspace{-0.4cm}\\
\end{center}
\caption{Particle-particle correlation function $g(r)$
plotted against the normalized center-center distance $r/2a$.
(Left) Disordered systems -- the arrows indicate the values at 
contact from Eq.\ (\protect\ref{eq:g2a}). (Right) Ordered systems with
different axis scaling.}
\label{fig:corr_g}
\end{figure}

\noindent
In {\bf monodisperse systems}, the particle-particle pair correlation function 
at contact, 
\begin{equation}
g_{2a}(\nu) = \frac{1-7\nu/16}{(1-\nu)^2} ~,
\label{eq:g2a}
\end{equation}
depends on the volume fraction only
\cite{chapman60.Luding,jenkins85b.Luding,hansen86,luding98f,sunthar99}.
The particle-particle correlation function is obtained from the 
simulations by averaging over $M$ snapshots with $N$ particles,
normalized to the value $g(r\gg 2a)=1$ for long distances in large
systems, so that
\begin{equation}
g(r)=\frac{1}{M} \sum_{m=1}^M  \frac{2 V}{N (N-1)} 
        \frac{1}{V_r} \sum_{i=1}^N \sum_{j=1}^{i-1}
        \theta[r_{ij}-r] \theta[r+\Delta r - r_{ij}] ~,
\label{eq:grnu}
\end{equation}
with $r_{ij}=|{\mathbf r}_i-{\mathbf r}_j|$, and
where the two $\theta$ functions select all particle pairs $(i,j)$ with 
distance between $r$ and $r+\Delta r$.  The weight $N(N-1)/2$ accounts
for all pairs summed over, and the term $V_r=\pi (2r+\Delta r) \Delta r$
is the volume (area) of a ring with inner radius $r$ and width $\Delta r$.
In Fig.\ \ref{fig:corr_g}, simulation results from set B (see table\
\ref{tab:simpar}) with different $\nu$ are presented.  Typical values 
used for the averages are e.g.~$M=50$, and $\Delta r = a/10$. 
Besides fluctuations, the values at contact nicely agree with the 
theoretical predictions from Eq.\ (\ref{eq:g2a}), as indicated by the arrows --
as long as the system is disordered (left panel in Fig.\ \ref{fig:corr_g}). 
In a more ordered system (right panel in Fig.\ \ref{fig:corr_g}),
$g_{2a}(\nu)$ is not a good estimate. Instead,
one obtains a long range order with peaks at $r/2a=1,\sqrt{3},2$, ... ,
indicating the triangular lattice structure of the assembly.

\begin{figure}[ht]
\begin{center}
\hspace{-.4cm}
\psfig{file=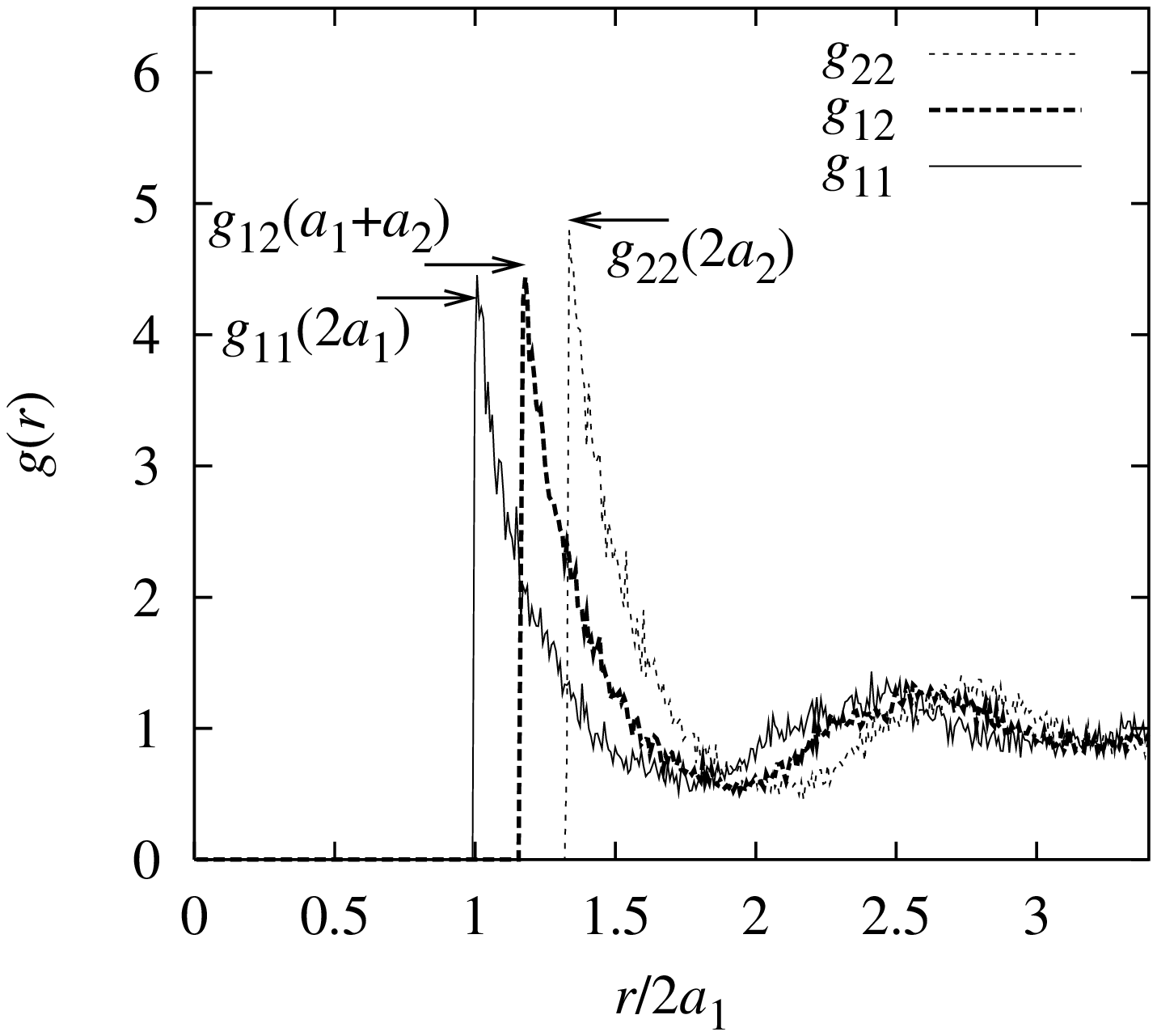,width=5.6cm,angle=-90} \hspace{-.8cm}
\psfig{file=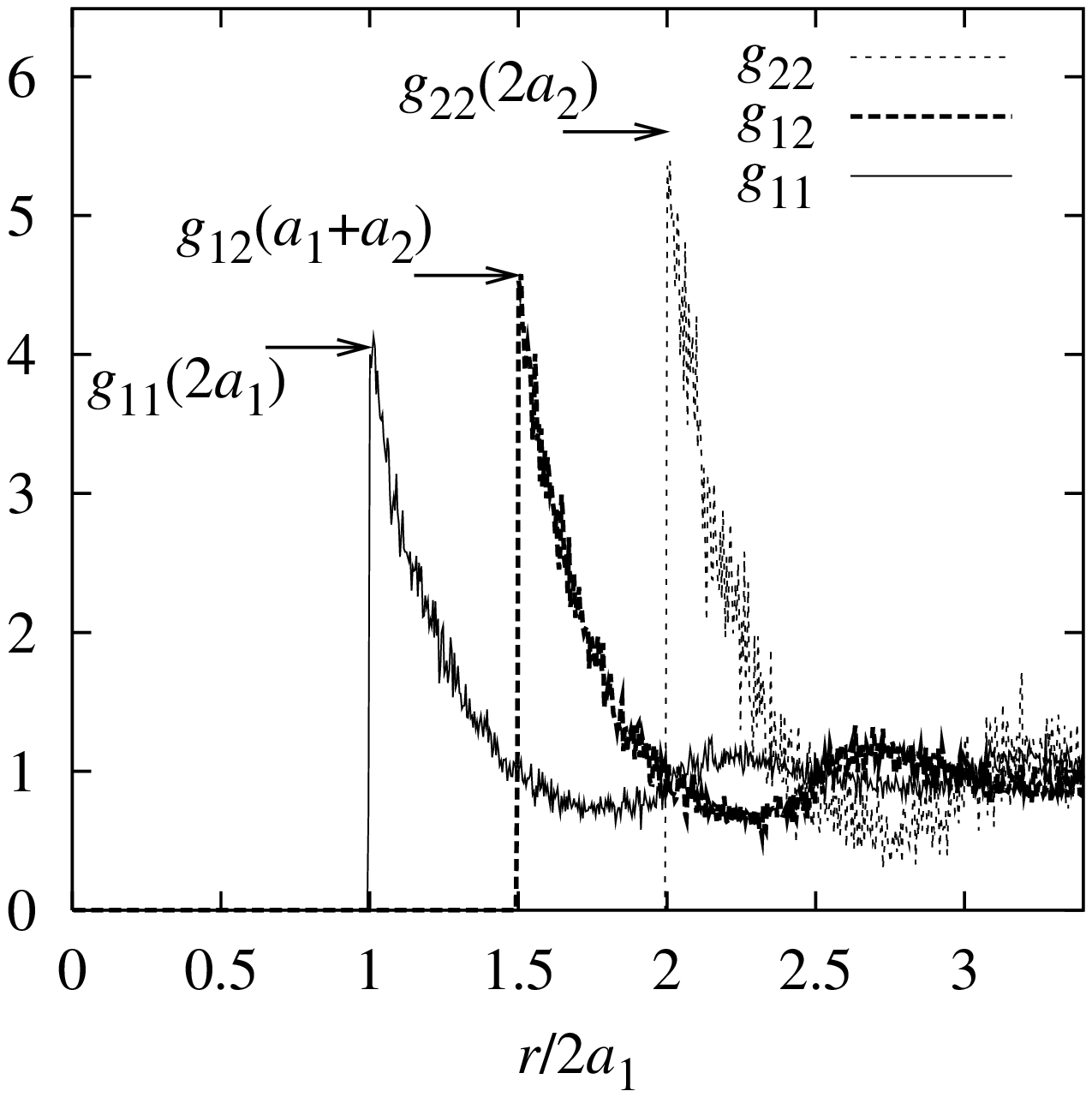,width=5.6cm,angle=-90}\\
~\vspace{-0.4cm}\\
\end{center}
\caption{Particle-particle correlation function $g(r)$
plotted against the center-center distance normalized
by the radius $a_1$ of the smaller particles,
at a volume fraction of $\nu=0.60$. The arrows indicate
the values $g_{11}$,  $g_{12}$, and $g_{22}$
at contact.
(Left) Bidisperse simulation from set C with $R=3/4$, and
(Right) bidisperse simulation from set D1 with $R=1/2$.
}
\label{fig:corr_gAB}
\end{figure}
\noindent
For {\bf bidisperse situations}, the pair correlation functions for equal 
species $g_{\rm 11}$ and $g_{\rm 22}$ are obtained by replacing $N$ in 
the first particle-sum in Eq.\ (\ref{eq:grnu}) by
$N_{\rm 1}$ and $N_{\rm 2}$, respectively. For the correlation
function $g_{\rm 12}$, it is necessary to perform the first particle-sum
in Eq.\ (\ref{eq:grnu}) from $i=1$ to $N_{\rm 1}$, the second sum from
$j=1$ to $N_{\rm 2}$, and to replace the weight $N(N-1)/2$ by 
$N_{\rm 1} N_{\rm 2}$ (in order to account for all pairs of different kind).  
In Fig.\ \ref{fig:corr_gAB} simulation results 
from sets C and D1 for $\nu=0.6$ are compared to the analytical 
expressions Eqs.\ (90), (91), and (92) from Ref.\ \cite{jenkins87},
here expressed in terms of $A_{1,2}$, $R$, and $\nu$:
\begin{eqnarray}
\label{eq:g11}
g_{\rm 11}&=&\frac{1-\nu \left ( 1-\frac{9}{16} \frac{A_1}{A_2} \right )}{(1-\nu)^2} ~,\\
\label{eq:g12}
g_{\rm 22}&=&\frac{1-\nu \left ( 1-\frac{9}{16R} \frac{A_1}{A_2} \right )}{(1-\nu)^2} ~,
{\rm ~~and}\\
\label{eq:g22}
g_{\rm 12}&=&\frac{1-\nu \left ( 1-\frac{9}{8(1+R)} \frac{A_1}{A_2} \right )}{(1-\nu)^2} ~.
\end{eqnarray}
Note that all $g_{ij}$ are identical to $g_{2a}(\nu)$ in the monodisperse
case with $R=1$ and $A_1 = A_2 = 1$.  The parameters used for averaging were 
$M=50$ and $\Delta r = a_1/118$ (Left) and $\Delta r = a_1/64$ (Right).
A finer binning leads to stronger fluctuations, a rougher binning 
does not resolve the values at contact, however, within the statistical 
error, the agreement between theoretical predictions and numerical results
is reasonable.

\begin{figure}[t]
\begin{center}
\hspace{-.4cm}
\psfig{file=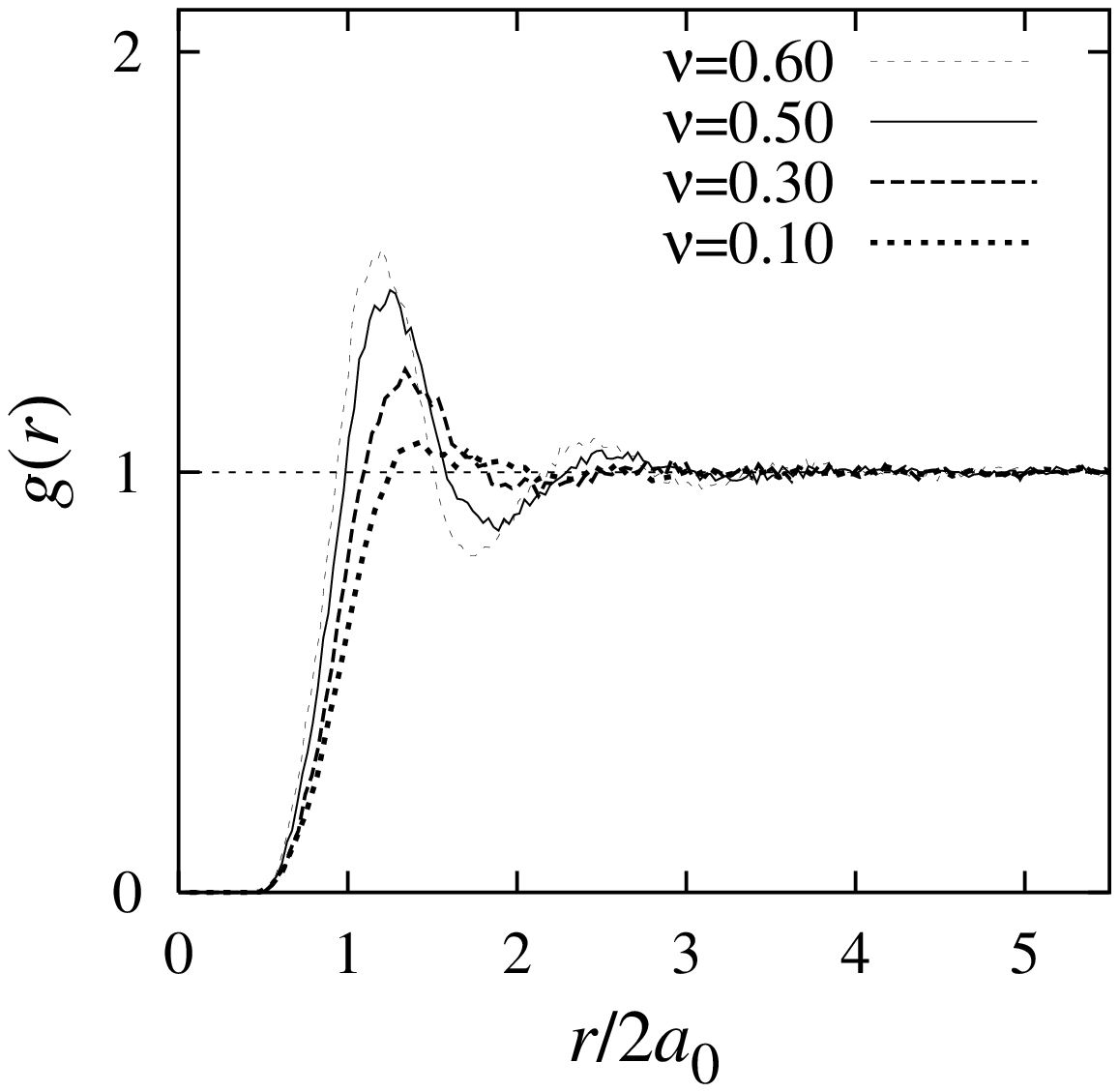,width=5.6cm,angle=-90} \hspace{-.8cm}
\psfig{file=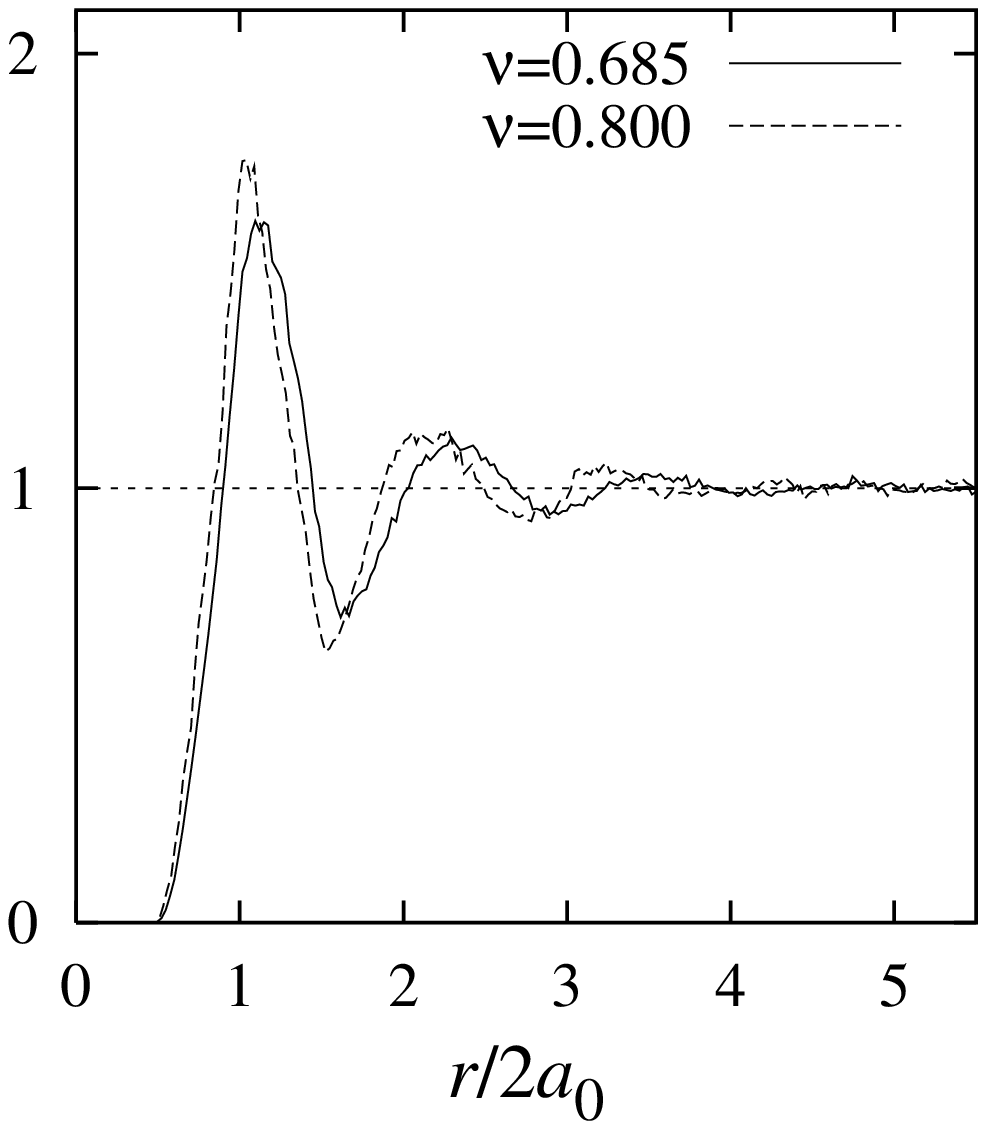,width=5.6cm,angle=-90}\\
~\vspace{-0.4cm}\\
\end{center}
\caption{Particle-particle correlation function $g(r)$
plotted against the normalized center-center distance $r/2a_0$.
(Left) Low density systems and (Right) high density systems with
the same axis scaling.}
\label{fig:corr_p}
\end{figure}
\noindent
Finally, in Fig.\ \ref{fig:corr_p}, particle correlation functions from 
{\bf polydisperse simulations} (set F1) are presented. Due to the broad
and continuous size distribution function, $g(r)$ is a smooth
function with much less variety in magnitude than in the mono-
and polydisperse situations discussed above. It more resembles
the distribution function of a gas or liquid with a smooth interaction 
potential \cite{hansen86}.

\subsection{Collision rates and energy dissipation}

In order to estimate the rate of change of energy in the
system $\dot T = t_E^{-1} \Delta T$, the collision frequency
$t_E^{-1}$ is needed. Rather than going into details concerning
the calculation of $t_E^{-1}$, we will simply use the Enskog
collision rate \cite{ziman79,hansen86,luding98f} for identical particles,
\begin{equation}
t_E^{-1} = \frac{4aN}{V} \sqrt{\pi} g_{2a}(\nu) \sqrt{T/m} ~,
\label{eq:crate}
\end{equation}
and, equivalently, the inter-species collision rates
\begin{equation}
t_{ij}^{-1} = \frac{ v^{\rm rel}_{ij} }{ \lambda_{ij} }
            = \frac{2 a_{ij} N_j}{V} \sqrt{\pi} g_{ij} \sqrt{T/(2m_{ij})} ~,
\label{eq:crateij}
\end{equation}
where all rates give the number of collisions of a particle per unit time, 
with $a_{ij}=a_i+a_j$. The temperature is here assumed to be 
independent of the particle species for the sake of simplicity.
This is approximately true in the systems examined below, provided they
stay rather homogeneous, but it is not true in general,
since the cooling rates depend on the species.

In Eq.\ (\ref{eq:crateij}), the term $\sqrt{T/(m_{ij})}$ 
is proportional to the mean relative velocity $v^{\rm rel}_{ij}$
of a pair ($i$, $j$), so that the remainder $t_{ij}^{-1}/v^{\rm rel}_{ij}$ 
can be seen as a measure for the inverse interspecies mean free path 
$\lambda_{ij}$.  The mean collision rate in the system is 
\begin{equation}
{\cal T}^{-1}_{\rm mix}  =  \sum_{i,j} n_i t_{ij}^{-1} 
                    =  \frac{4 a_1 N}{V} \sqrt{\pi} \sqrt{T/m_1} 
                        \sum_{i,j} n_i n_j g_{ij} c_{ij}  ~,
\label{eq:Tmix}
\end{equation}
with $c_{11}=1$, $c_{12}=(1+R)/(2R) \sqrt{(1+R^3)/2}$, and 
$c_{22}=\sqrt{R}$. Note that $c_{12}$ and $c_{22}$ depend on
mass and density of the different species.
The mean collision rate was tested for the monodisperse and bidisperse
situations and showed the same quality of agreement as the pressure,
which will be discussed in the next subsection. Therefore, we do
not present data of the collision rates here, but perform a detailed
numerical study of the mixture pressure below. However, we should remark
that the interspecies collision rates are of the same order of magnitude,
even if the species fluctuation velocities $v_i = \sqrt{T/m_i}$ strongly 
differ due to the differences in mass, also when the temperature 
$T=T_i=T_j=E/N$ is not species dependent.
This means that the mean free distance between collisions
compensates the speed; the distance traveled between collisions is 
proportional to the species velocity.
\begin{figure}[ht]
$\nu=0.30$ \hfill $\nu=0.60$\\ ~\vspace{-0.6cm}\\
\psfig{file=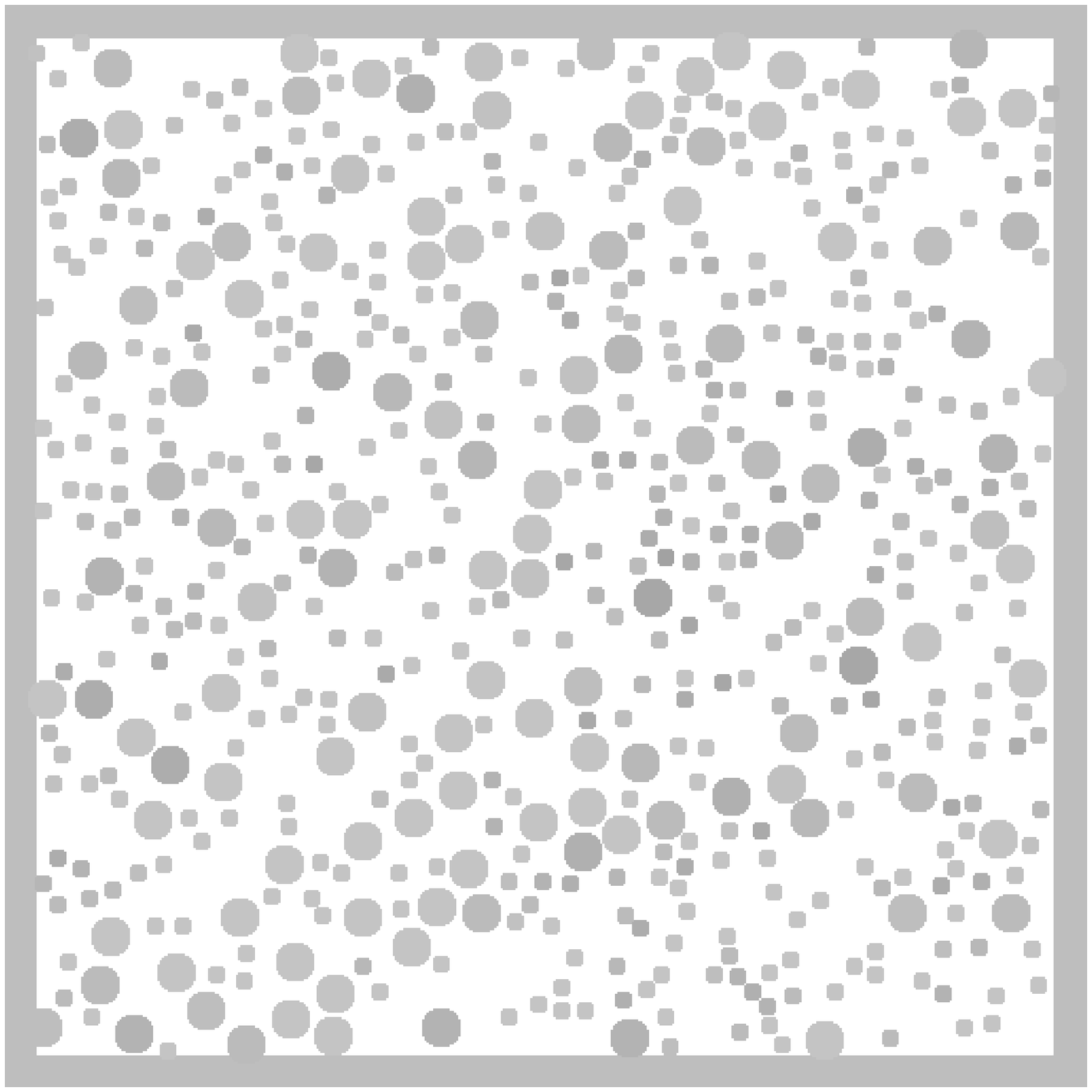,width=5.6cm,angle=-90}\hfill\psfig{file=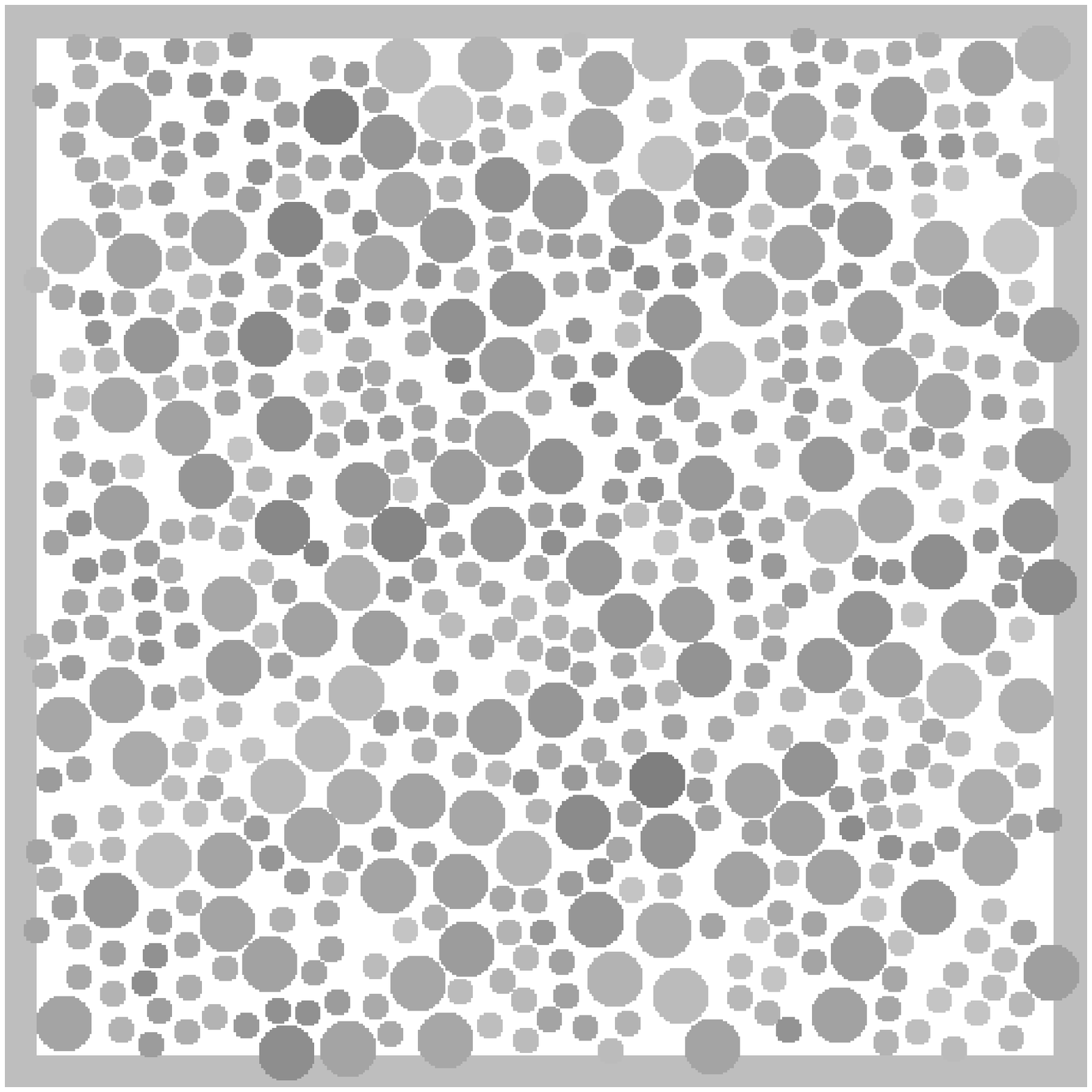,width=5.6cm,angle=-90}~\\ ~\vspace{0.1cm}\\
$\nu=0.70$ \hfill $\nu=0.85$\\ ~\vspace{-0.6cm}\\
\psfig{file=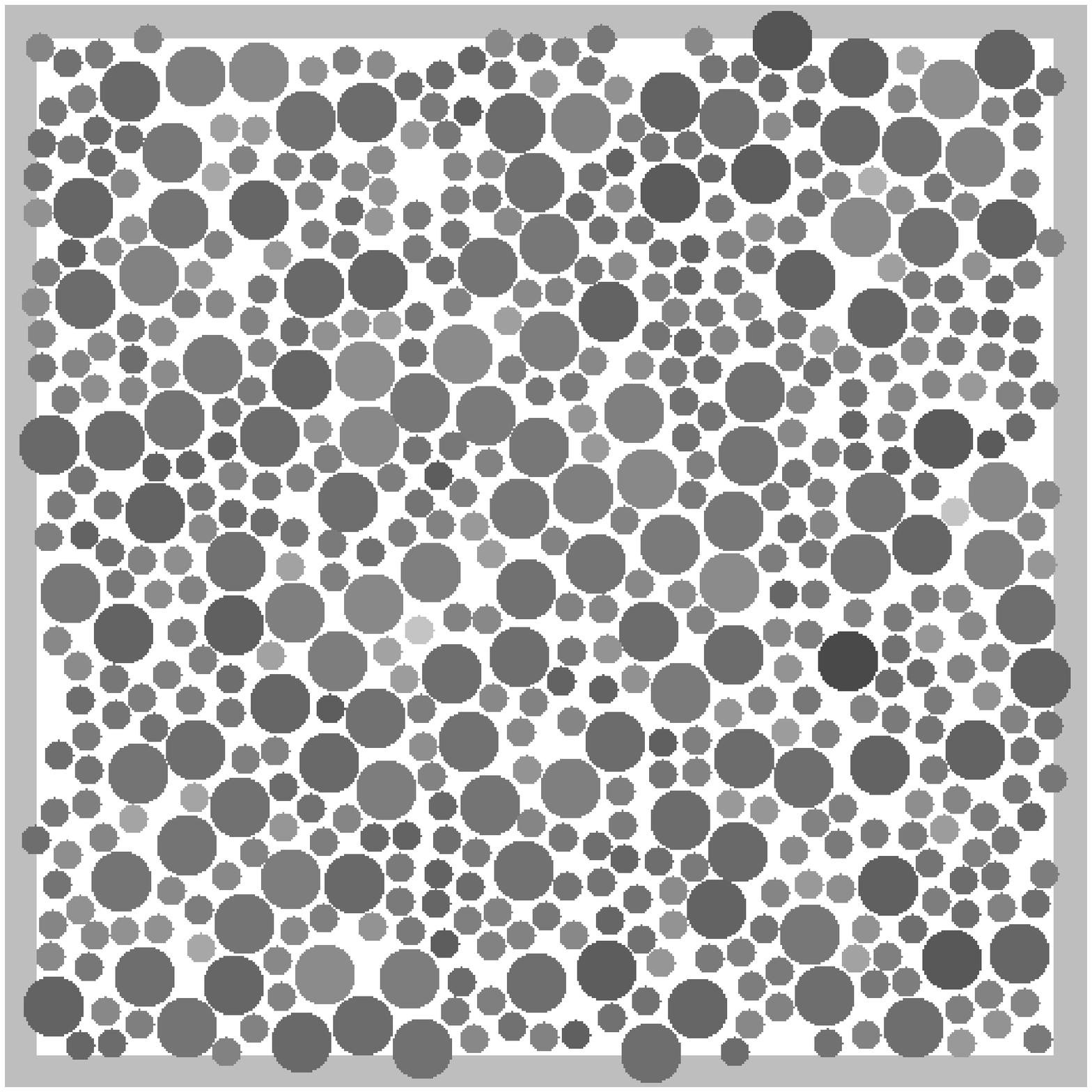,width=5.6cm,angle=-90}\hfill\psfig{file=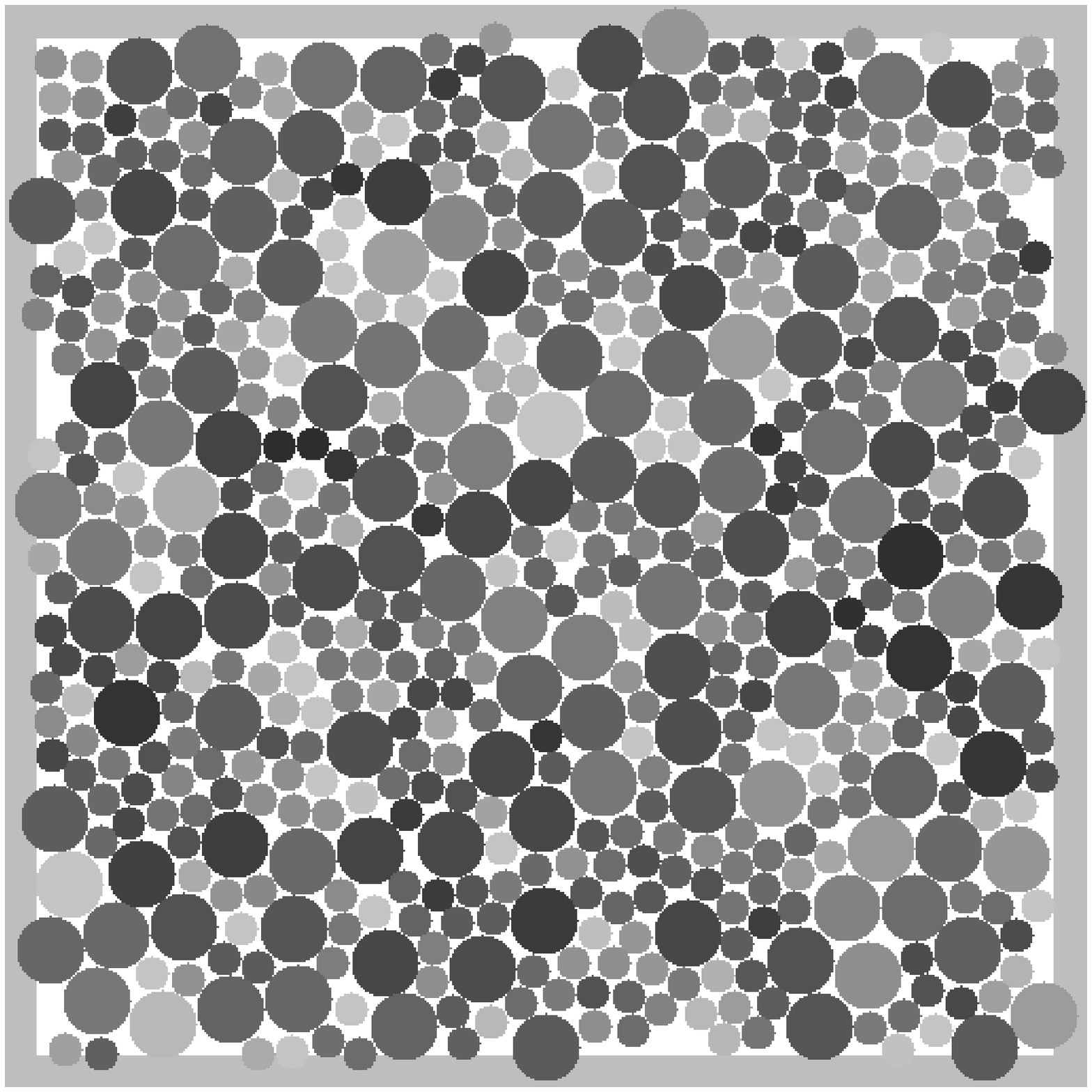,width=5.6cm,angle=-90}~\\
\caption{Snapshots from the bisdisperse simulations (set D1) for
different volume fractions $\nu$. The grey-scale denotes the collision
rate of the corresponding particle, the darkest particles had a collision
rate of $1500$\,s$^{-1}$, the lightest particles had a rate of
less than $275$\,s$^{-1}$. (These numbers have to be multiplied
with $2/5$ for $\nu=0.30$ and with $4$ for $\nu=0.85$, in order to
allow for a comparison of all pictures.)}
\label{fig:snap1}
\end{figure}
In Fig.\ \ref{fig:snap1} several snapshots from simulation set D1 are
presented. The grey-scale indicates the collision rates, dark particles
collide more frequently than light particles, and the collision rate 
increases with the density, but as discussed above, the collision rates
of the two species are comparable.

Knowing both the collision rates and the energy loss per collision 
\begin{equation}
\Delta T_{ij} = - m_{ij} \frac{1-r^2}{2} (v^{\rm rel}_{ij})^2
              = - \frac{1-r^2}{2} T ~,
\end{equation}
it is straightforward to compute the decay of energy as a function of time
\begin{equation}
\frac{dT}{dt} = \sum_{i,j} n_i t_{ij}^{-1} \Delta E_{ij} 
              = - \frac{1-r^2}{2} {\cal T}_{\rm mix}^{-1}(t) T(t)~,
\end{equation}
where both $T(t)$ and ${\cal T}_{\rm mix}^{-1}(t) \propto \sqrt{T(t)}$
depend on time. The differential equation is easily solved and one gets 
the scaled temperature
\begin{eqnarray}
\frac{T}{T_0} = \left ( 
                1+\frac{1-r^2}{4} {\cal T}_{\rm mix}^{-1}(0) \,t 
                \right )^{-2} ~,
\label{eq:Tskal}
\end{eqnarray}
identical to the solution of the homogeneous cooling state 
of monodisperse disks \cite{luding98d}, where $t_E^{-1}(0)$ is 
replaced by ${\cal T}_{\rm mix}^{-1}(0)$ from Eq.\ (\ref{eq:Tmix}).

\begin{figure}[ht]
\begin{center}
\hspace{-3.4cm}
\psfig{file=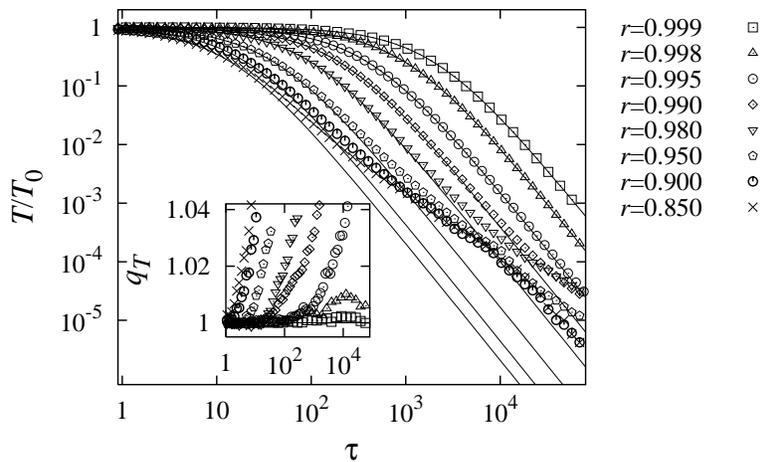,width=6.7cm,angle=-90}
\end{center}
\caption{Dimensionless temperature $T/T_0$ plotted against rescaled time
$\tau =  {\cal T}_{\rm mix}^{-1}(0) t$ for different $r$, with double 
logarithmic axis. The symbols are simulation results from set D2, the 
solid lines correspond to Eq.\ (\protect\ref{eq:Tskal}). 
In the inset the quality factor $q_T=T_{\rm sim}/
T_{\rm theory}$ is plotted against the time $\tau$.
}
\label{fig:Tskal}
\end{figure}

In Fig.\ \ref{fig:Tskal}, simulations from set D2 are presented
for different values of $r < 1$. The agreement between simulations
and Eq.\ (\ref{eq:Tskal}) is perfect for short times.  With decreasing
$r$, i.e.~increasing dissipation, the deviations from the theory
occur earlier due to the break-down of the homogeneity and
the related simplifying assumptions of molecular chaos and Gaussian
velocity distributions \cite{luding98e,noije98c}.

\subsection{Stress and the equation of state}

The stress tensor, defined for a test-volume $V$, has two
contributions, one from the collisions and the other from the
translational motion of the particles. Using $a$ and $b$
as indices for the cartesian coordinates one has the components
of the stress tensor (where the  sign is convention)
\begin{equation}
\sigma^{ab} = \frac{1}{V}
   \left [
      \sum_i m_i v_i^a v_i^b
      - \frac{1}{\Delta t} \sum_n \sum_{j=1,2} \Delta p_j^a \ell_j^b  
   \right ] ~,
\end{equation}
with $\ell_j^b$, the components of the vector from the center of mass
of the two colliding particles $j$ to their contact points at collision 
$n$, where the momentum $\Delta p_j^a$ is exchanged.
The sum in the left term runs over all particles $i$, the
first sum in the right term runs over all collisions $n$ occuring in 
the time-interval $\Delta t$, and the second sum in the right term
concerns the collision partners of collision $n$ -- in any case the
corresponding particles must be within the averaging volume $V$ 
\cite{goddard86,emeriault97,luding98e,luding98f}.
Note that the results may depend on the choice of $\Delta t$ 
and $V$ \cite{goldhirsch98}, however, a discussion of different
averaging procedures and parameters is far from the scope of this
study.

\begin{figure}[ht]
\begin{center}
\psfig{file=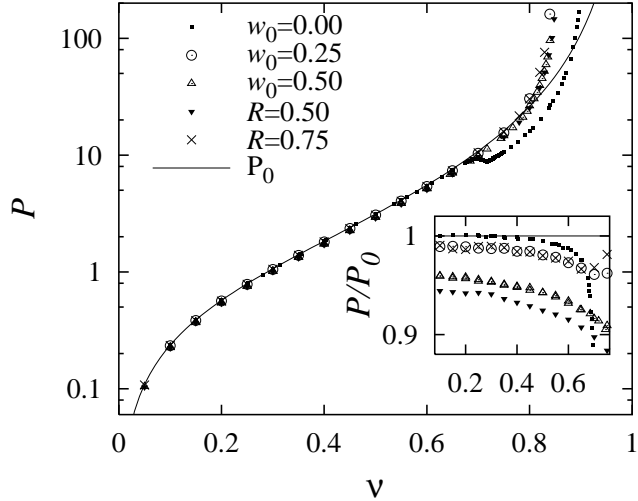,height=9cm,angle=-90}
\end{center}
\caption{Dimensionless pressure $P=pV/E-1$ plotted against the
volume fraction $\nu$ for different particle size-distribution
functions. The bidisperse situation is identified by the size
ratio $R$. The solid line corresponds to $P_0$.  In the inset
the quality factor $P/P_0$ is plotted for the same data. The wiggle
of the $w_0=0$ data is discussed in more detail below.}
\label{fig:pressure}
\end{figure}

The mean pressure $p = (\sigma_1 + \sigma_2)/2~$, with the eigenvalues
$\sigma_1$ and $\sigma_2$ of the stress tensor, can be obtained from 
simulations with rigid, elastic particles ($r=1$) and different volume 
fractions $\nu$ \cite{luding98f,sunthar99}. 
The dimensionless reduced pressure from simulations
agrees perfectly with the theoretical prediction \cite{jenkins85b.Luding}
\begin{equation}
P_0 = P V / E - 1 = 2 \nu g_{2a}(\nu) ~,
\label{eq:P0}
\end{equation}
with the total energy $E = (1/2) \Sigma_{i=1}^N m_i {\vec v}_i^2$.
In Fig.\ \ref{fig:pressure} the simulation results for
the dimensionless pressure $P = pV/E-1$ are compared to
the kinetic theory result $P_0 = 2 \nu g_{2a}(\nu)$ \cite{luding98f}. 

When plotting $P$ against the volume fraction $\nu$ with a 
logarithmic vertical axis, the results for the different 
simulations can not be distinguished for $\nu < 0.7$. 
In the monodisperse system, we obtain crystallization around 
$\nu=0.7$, and the data clearly deviate from $P_0$, i.e.~the 
pressure is strongly reduced due to crystallization and, thus,
enhanced free volume. The monodisperse data diverge at the maximum
packing fraction $\nu^{\rm mono}_{\rm max}=\pi/(2\sqrt{3})$ in 2D.
Note that one has to choose the system size such that a triangular
lattice fits perfectly in the system, i.e. $L_y/L_x=\sqrt{3} h/2w$
with integer $h$ and $w$ -- otherwise the maximum volume fraction
is smaller.
All other simulations are close to $P_0$ up to $\nu \approx 0.8$,
where they begin to diverge; the bidisperse data with $R=1/2$, for 
example, diverge at $\nu^{\rm bi}_{\rm max} \approx 0.858$. The maximum 
packing fraction is smaller for the polydisperse size distributions
used here and the crystallization, i.e.~the pressure drop, does not
occur for polydisperse packings with $w_0~_\approx^{>}~0.15$ 
(the data which lead to this approximate result are not shown).
Since the logarithmic axis hides small deviations, we plot also
the quality factor $P/P_0$ of the data in the inset. In this 
representation, values of unity mean perfect agreement, while
smaller values correspond to an overestimation of the data by
a factor of $P_0/P$ when $P_0$ would be used instead of the 
simulation results $P$. The deviations increase with increasing 
width of $w(a)$ and with increasing volume fraction. Note that
there exists a deviation already for small $\nu$. 

\begin{figure}[ht]
\begin{center}
$~~~~~~~$\psfig{file=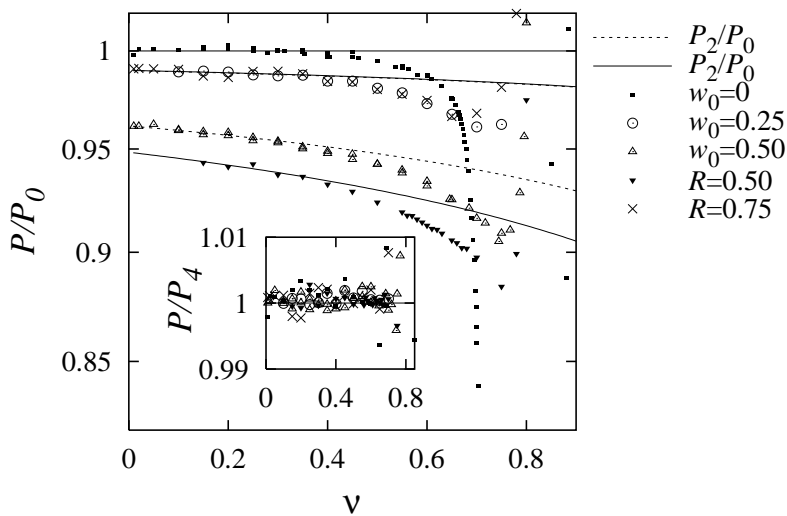,height=19cm,angle=-90}
~\vspace{-0.3cm}\\
\end{center}
\caption{Quality factor $P/P_0$ from the inset of
Fig.\ \protect{\ref{fig:pressure}}. The lines give 
$P_2(\nu)/P_0$ from Eq.\ (\protect{\ref{eq:p2}}). In the 
inset, the simulation data for $P$ are rescaled by $P_4(\nu)$
from Eq.\ (\protect{\ref{eq:p4}}).}
\label{fig:p2}
\end{figure}

A more elaborate calculation in the style of Jenkins and Mancini,
see Eq.\ (60) in \cite{jenkins87}, leads to the partial translational
pressures $p^t_i = n_i E / V$ for species $i$ and to the collisional 
pressures
$p^c_{ij} 
         = \pi N_i N_j g_{ij} a_{ij}^2 (1+r_{ij}) T / (4 V^2)$
with the particle correlation functions from Eqs.\ (\ref{eq:g11})-(\ref{eq:g22})
evaluated at contact, and $a_{ij}=a_i+a_j$. In the simulations from
Fig.\ \ref{fig:pressure}, the inter-species restitution coefficients are
equal and elasticity is assumed, $r=r_{11}=r_{12}=r_{22}=1$.
Note that the species temperatures are equal, so that the corresponding 
correction term can be dropped. Thus, the global pressure in the mixture is 
\begin{eqnarray}
p^{\rm m}
  & = & p^t_1 + p^t_2 + p^c_{11} + 2 p^c_{12} + p^c_{22}   \nonumber \\
  & = & \frac{E}{V} \left [ 1 + (1+r) \frac{\nu}{A_2 a_{11}^2}  
                      (   g_{11} a_{11}^2 n_1^2 
                      + 2 g_{12} a_{12}^2 n_1 n_2
                      +   g_{22} a_{22}^2 n_2^2 )   \right ]  \nonumber
~\\
  & = & \frac{E}{V} \left [ 1 + {(1+r)\nu} g_{\cal A}(\nu) \right ] 
\label{eq:ptotal}
                   ~.   
\end{eqnarray}
Assuming a monodisperse system as a test case, i.e.~ inserting 
$R=A_1=A_2=1$, into Eq.\ (\ref{eq:ptotal}), leads to the monodisperse solution
$p^{\rm m} V/ E-1=P_0$, as expected.  The effective correlation
function $g_{\cal A}(\nu)$ can be expressed in terms of the width-correction
${\cal A}$ of the size distribution so that 
\begin{equation}
g_{\cal A}(\nu)
 = \frac{ \left (1+{\cal A}\right ) -\nu \left (1-{\cal A} /8 \right )}
        {2 (1-\nu)^2} ~,
\label{eq:geff}
\end{equation}
with ${\cal A} = {\langle a \rangle^2}/{\langle a^2 \rangle}$.
Note that ${\cal A}$ is well defined for any size distribution
function, so that Eq.\ (\ref{eq:geff}) can also be applied to
polydisperse situations.
In the limit of small volume fraction $\nu \rightarrow 0$, one can
estimate the normalized pressure by 
\begin{equation}
P_1 = (1+r) \nu g_{2a}(\nu) \frac{1+{\cal A}}{2} ~,
\end{equation}
as proposed by Zhang et al.\ \cite{zhang99}, 
when disregarding the dependence of $g(r)$ on the types of the 
collision partners.
The values of $P/P_1$ in the limit $\nu \rightarrow 0$ agree very
well with the simulations.  Using the effective particle correlations,
one can define 
\begin{equation}
P_2(\nu) = \frac{ p^{\rm m} V }{ E } - 1 = (1+r) \nu g_{\cal A}(\nu) ~,
\label{eq:p2}
\end{equation}
and compare the resulting expected reduced pressure with the
simulation results from Fig.\ \ref{fig:pressure}. An almost perfect 
agreement between $P$ and $P_2(\nu)$ is obtained for $\nu < 0.4$
and even for larger $\nu \approx 0.65$, the difference is always less 
than about two percent, and, the quality factors for {\em all} simulations
collapse.  Note that the quality is perfect (within less than 0.5 percent
for all $\nu < 0.65$) if $P_2(\nu)$ is multiplied by the empirical 
function $1-\nu^4/10$, as fitted to the quality factor $P/P_2$. 
Thus, based on our simulation results, we propose the corrected,
nondimensional mixture pressure
\begin{equation}
P_4(\nu) = \frac{ p^{\rm m} V }{ E } - 1 
         = (1+r) \nu g_{\cal A}(\nu) \left [ 1- a_g \nu^4 \right ]~,
\label{eq:p4}
\end{equation}
with the empirical constant $a_g \approx 0.1$, for the pressure for all
$\nu < 0.65$. For larger $\nu$ the excluded volume effect becomes more
and more important, leading to a divergence of $P/P_4$. Furthermore,
in the high density regime, the behavior is strongly dependent on the 
width of the size distribution function, see Fig.\ \ref{fig:p2}. 

\subsection{Accounting for the dense, ordered phase}

\begin{figure}[htb]
  \center{
     \hspace{-0.5cm}
     ~\psfig{file=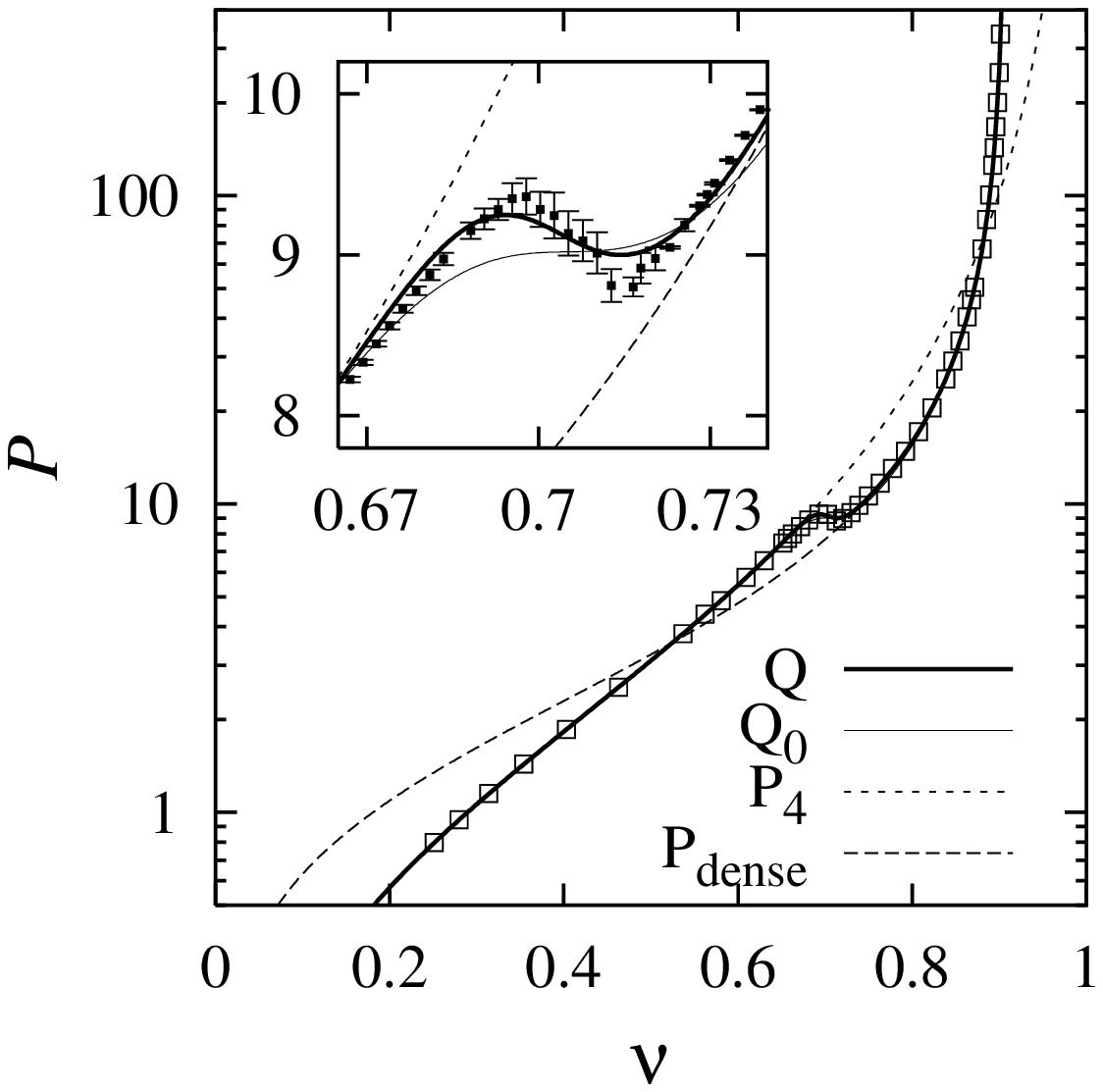,height=6.0cm,angle=-90,clip=} 
     \hspace{-0.5cm}
     ~\psfig{file=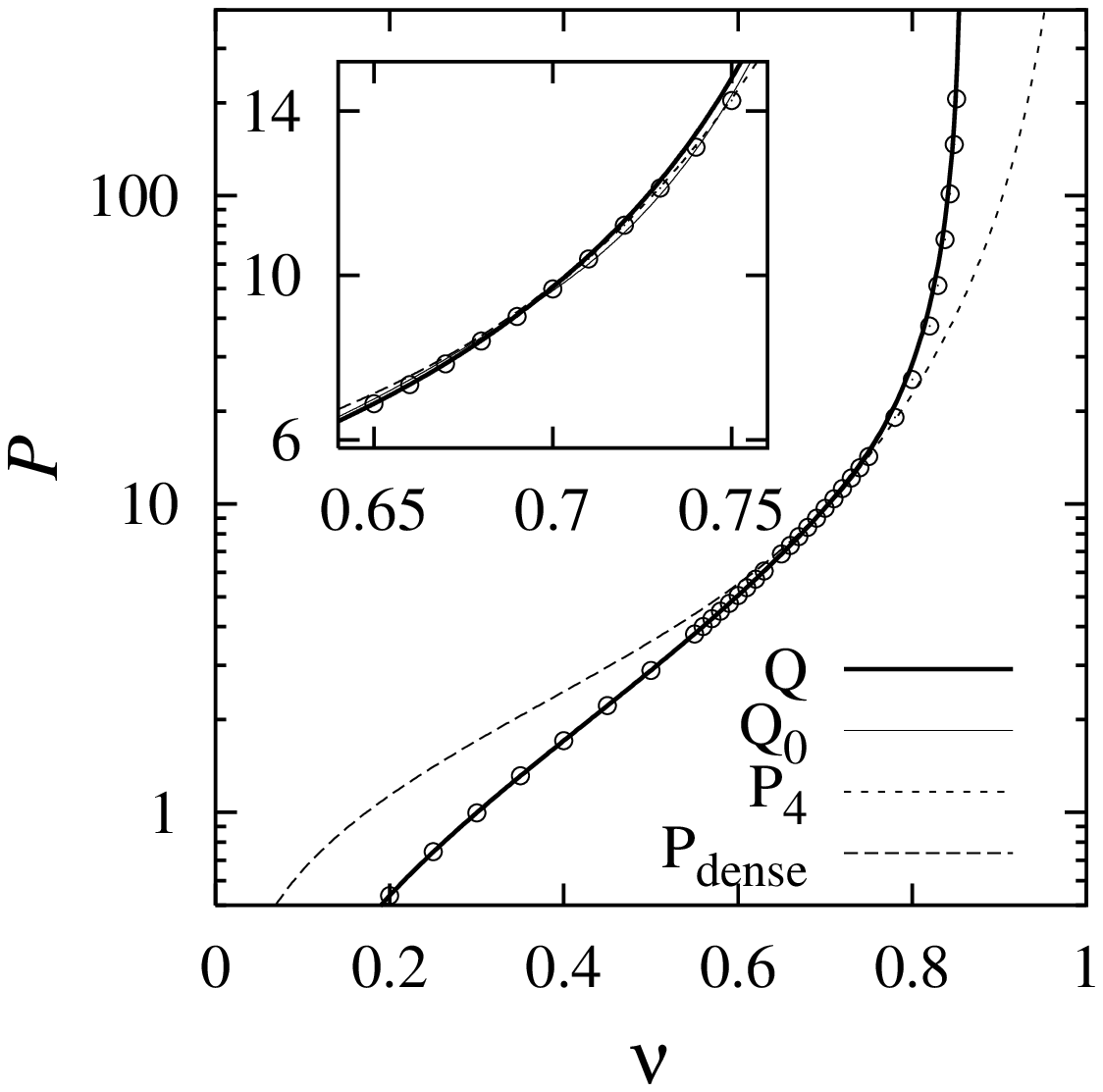,height=6.0cm,angle=-90,clip=}
    }
  \caption{Dimensionless pressure $P$ from set B simulations (symbols)
    plotted against the volume fraction $\nu$. (Left) The dashed lines are 
    $P_4$ from Eq.\ (\ref{eq:p4}) and $P_{\rm dense}$ from Eq.\ (\ref{eq:pdense}).
    The thick solid line is $Q$, the corrected global equation of state from
    Eq.\ (\ref{eq:pgeneral}) with the fit-parameters $a_g=0.1$, $a_d=0.340$, $a_p=1.09$, 
    $\nu_c=0.701$, $\nu_{\rm max}=0.9069$, and $m_0=0.00928$. The thin solid line 
    is $Q_0$ without corrections, i.e.~ $a_g=0$, $a_d=0$, and $m_0=0.0015$ and $\nu_c=0.7$,
    so that $Q_0=P_2+m(\nu) [P_{\rm fv} - P_2]$.  (Right)  Analogous data from
    set D1, where only $\nu_{\rm max}=0.858$ is different from
    the left panel.
    }
  \label{fig:pdense}
\end{figure}

The equation of state in the dense, ordered phase has been
calculated by means of a free volume theory 
\cite{kirkwood50,buehler51,wood52}, that leads in 2D to the 
reduced pressure $P_{\rm fv}=1/(\sqrt{\nu_{\rm max}/\nu}-1)$
with the maximum volume fraction $\nu_{\rm max}$. 
Based on our simulation results we propse the corrected high
density pressure
\begin{equation}
P_{\rm dense}=\frac{1}{\sqrt{\nu_{\rm max}/\nu}-1} 
              \left [ 1+a_d (\nu_{\rm max}-\nu)^{a_p} \right ]~,
\label{eq:pdense}
\end{equation}
where the term in brackets $[ \ldots ]$ is a fit function 
with $a_d=0.340$ and $a_p=1.09$. The special case $a_d=0$ leads to the 
theoretical result $P_{\rm fv}$.
In the left panel of Fig.\ \ref{fig:pdense}, data from set B (see table\ 
\ref{tab:simpar}) are presented, together with the ``low'' and ``high'' density
predictions $P_4$ and $P_{\rm dense}$, respectively (dashed and dotted lines).

To our knowledge, no theory exists, which combines the disordered and
the ordered regime. Therefore, we propose a global equation of
state
\begin{equation}
Q = P_4 + m(\nu) [P_{\rm dense} - P_4] ~,
\label{eq:pgeneral}
\end{equation}
with an empirical merging function
\begin{equation}
m(\nu) = \frac{1}{1+\exp\left (-(\nu-\nu_c)/m_0 \right )} ~
\end{equation}
which selects $P_4$ for $\nu \ll \nu_c$ and $P_{\rm dense}$ for
$\nu \gg \nu_c$ with the width of the transition $m_0$. In Fig.\
\ref{fig:pdense}, the fit parameters $\nu_c \approx 0.70$ and 
$m_0 \approx 0.009$ lead to qualitative and quantitative agreement 
between $Q$ (thick line) and the simulation results (symbols). 
However, also a simpler version $Q_0$ (thin line) without numerical 
corrections leads to reasonable agreement when $m_0=0.015$ is used, 
except for the transition region. The pressure drop when $\nu$ 
is increased above $\nu_c$ is qualitatively reproduced but no 
negative slope occurs. Due to the latter fact, the expression $Q_0$ 
allows for an easy numerical integration of $P$.  
We selected the parameters for $Q_0$ 
as a compromise between the quality of the fit on the one hand and 
the treatability of the function on the other hand. 

Remarkably, as one can see from Fig.\ \ref{fig:pdense} (Right), the 
dimensionless pressure $Q$ from Eq.\ (\ref{eq:pgeneral}) describes,
at least qualitatively, the behavior of the polydisperse simulations 
when $\nu_{\rm max}=0.858$ is used.  Note that the
pressure drop at the transition $\nu_c \approx 0.7$ from the low
density, disordered regime to the high density, ordered regime,
is almost non-existent, since $P_4$ and $P_{\rm dense}$ are almost
collapsing in this range of density.

\section{Pressure gradient due to gravity}
\label{sec:pgrad}

In an experiment on earth, usually gravity plays an important
role, it introduces a pressure gradient. Therefore, the density 
and pressure profiles of granular systems in equilibrium in the 
gravitational field are examined in the following.
Here, a horizontal wall at $z=0$ is introduced in a periodic two 
dimensional system of width $L=l_x/(2a)$, infinite height, and 
the gravitational acceleration $g=1$\,ms$^{-2}$.

\subsection{Density profile in dilute systems}

In the special case of low density, one can use the equation of state of
an ideal gas and express the pressure as a function of the energy density:
\begin{equation}
  \label{eq:ideal1}
  p = \frac{E}{V} = n T ~, 
\end{equation}
with the number density $n=N/V=n(z)=\nu(z)/(\pi a^2)$
and the ``granular temperature'' $T=E/N$ in two dimensions.

The gradient of pressure ${dp}/{dz}$ compensates for the weight 
$n m g L dz$ of the particles in a layer of height $dz$, so that
\begin{equation}
  \label{eq:ideal2}
  \frac{dp}{dz} = \frac{dp}{d\nu}\frac{d\nu}{dz} 
                = -n m g 
\end{equation}
Separation of variables and the assumption of a constant temperature
leads to the density profile for an ideal gas
\begin{equation}
  \label{eq:ideal4}
  \nu(z) = \nu_0 \exp \left( -\frac{mg(z-z_0)}{T} \right) 
{\rm ~~~ or ~~~ }
  z(\nu) = z_0+{z_T} \ln\frac{\nu_0}{\nu} ~,
\end{equation}
with $\nu<\nu_0$ and $z_T=T/(mg)$. In a system with a constant 
particle number $N$, one has
\begin{equation}
  \label{eq:nu0def}
  N \stackrel{!}{=} 
  \frac{L}{\pi a^2}\int_{z_0}^{\infty} \nu(z) dz =
  \frac{L}{\pi a^2}\int_0^{\nu_0} \left ( z(\nu)-z_0 \right ) d\nu ~.
\end{equation}
Eq.\ (\ref{eq:nu0def}) allows to determine analytically the volume fraction 
$\nu_d$ at the bottom $z_0$, in the dilute limit, by integration of $z(\nu)$
\begin{equation}
  \label{eq:ideal7}
  \nu_d = \frac{N \pi a^2 m g}{TL} 
        = \frac{N \pi a^2}{z_T L} ~,
\end{equation}
defined here for later use.

This case can be extended to dilute and weakly dissipative systems, 
since the temperature is almost constant except for the bottom 
boundary layer \cite{sunthar99,eggers99b}. In the following we rather 
extend it to arbitrary density, but keep $r=1$.

\subsection{Density profile for a monodisperse hard sphere gas}

In the dense case, Eq.\ (\ref{eq:ideal1}) is modified to
\begin{equation}
  \label{eq:pd}
  p = n T \left [ 1 + 2 \nu g_{2a}(\nu) \right ] ~, 
\end{equation}
using Eq.\ (\ref{eq:P0}) with $r=1$, and inserting Eq.\ (\ref{eq:pd})
into Eq.\ (\ref{eq:ideal2}) leads to
\begin{equation}
  \label{eq:fgrav}
  \pi a^2 \frac{dp}{dz} = \frac{d}{d\nu} \left [ 
                              \nu T (1+2 \nu g_{2a}(\nu) ) \right ] 
                          \frac{d\nu}{dz} 
                        = - \nu m g ~.
\end{equation}
Assuming again that $T$ is constant, one gets
\begin{equation}
  \label{eq:grav2}
  \frac{d}{d\nu}[\ldots]  
  = T \, \left\{ 1 + \frac{\partial}{\partial \nu} \left(
  2 \nu^2 g_{2a}(\nu) \right) \right\}
  = T \, \frac{8+ 8\nu+ 3\nu^2- \nu^3}{8(1-\nu)^3}  ~,
\end{equation}
which, inserted in Eq.\ (\ref{eq:fgrav}), allows integration 
from $\nu_0$ to $\nu$ and from $z_0$ to $z$:
\begin{equation}
  \label{eq:grav3}
  \int^{\nu}_{\nu_0} \left\{
    \frac{8}{\nu'}+\frac{7}{1-\nu'}+\frac{7}{(1-\nu')^2}+\frac{18}{(1-\nu')^3}
  \right\} d\nu'
  = - \frac{8 m g}{T} \int_{z_0}^{z} dz' ~,
\end{equation}
and leads to an implicit definition of $\nu(z)$:
\begin{equation}
  \label{eq:grav3b}
  \left[ \ln \nu' - \frac{7}{8} \ln (1-\nu') + 2 g_{2a}(\nu') 
  \right]_{\nu_0}^{\nu(z)}
  = -\frac{z-z_0}{z_T} ~.
\end{equation}
We express $z$ as a function of the volume fraction
\begin{equation}
  \label{eq:znu}
  \frac{z(\nu)-z_0}{z_T} = 
    \ln\frac{\nu_0}{\nu}-\frac{7}{8}\ln\frac{1-\nu_0}{1-\nu}+
     2 g_{2a}(\nu_0) - 2 g_{2a}(\nu) ~,
\end{equation}
with the unknown volume fraction $\nu_0$ at $z_0$, which,
however, is determined using Eq.\ (\ref{eq:nu0def}):
\begin{equation}
 \frac{N_0 \pi a^2}{z_T L} = \nu_d \stackrel{!}{=}  
   \int_0^{\nu_0} \frac{z(\nu)-z_0}{z_T} d\nu 
 = {\nu_0} \, \frac{8+\nu_0^2}{8(1-\nu_0)^2} ~,
\end{equation}
where $N_0$ is the number of particles above a given height
$z_0$. Only if $z_0=0$, one has $N_0=N$.
This leads to a third order polynomial for $\nu_0$, 
\begin{equation}
  \label{eq:nu0}
  \nu_0^3 -8 \nu_d \nu_0^2 +(16 \nu_d+8)\nu_0 - 8 \nu_d = 0 ~,
\end{equation}
which can be solved analytically \cite{bronstein79}, and always has 
at least one real solution. Note that the function $g_{2a}(\nu)$ is
wrong at high densities $\nu > \nu_c$, so that also the pressure is
not correct for high densities. This fact is discussed also
by D. Hong \cite{HongHere}, who performed the three dimensional 
calculations analogous to our 2D calculus in this section.

\subsection{Comparison with simulations}

\begin{figure}[tb]
  \center{
     ~\psfig{file=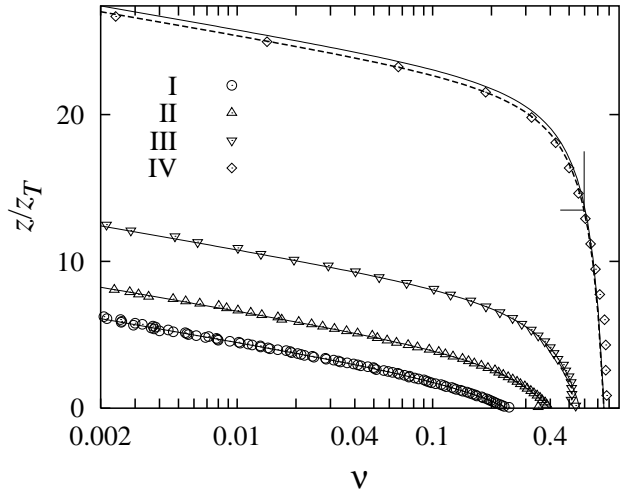,height=8.8cm,angle=-90,clip=}
    }
  \caption{Rescaled height $z/z_T$ plotted against the volume 
    fraction $\nu$ from four different simulations, see table\ 
    \protect\ref{tab:simgpara},
    with logarithmic horizontal axis. Symbols are simulation data
    and the lines correspond to Eq.\ (\protect\ref{eq:znu}) with the 
    respective value of $\nu_0$.
    }
  \label{fig:grav1}
\end{figure}

In this subsection, the theoretical density profile in Eq.\ (\ref{eq:znu}),
with the parameter $\nu_0$ determined via Eq.\ (\ref{eq:nu0}), is compared
to numerical simulations with the parameters as specified in table\
\ref{tab:simgpara}. In Fig.\ \ref{fig:grav1}, the rescaled height $z/z_T$ 
is plotted against the volume fraction $\nu$, according to Eq.\ (\ref{eq:znu}).
Note that even when the simulation parameters are rather arbitrary, 
the data follow a master-curve from $\nu=0$ to $\nu=\nu_0$ (or equivalently
from $z=\infty$ to $z=0$) only shifted vertically such that $z(\nu_0)=0$.
The agreement between simulation and theory is almost perfect, except for
simulation IV where densities above $\nu \approx 0.65$ are observed, i.e.
above the limit of validity of the equation of state.  Therefore, the
numerical values of $z/z_T$ are systematically smaller than the theoretical
line obtained for $\nu_0$ from table\ \ref{tab:simgpara}.  

\begin{table}[htb]
\begin{center}
\begin{tabular}{|r|l|l|l|l|l|l|}
\hline
    & $N$  & $L/(2a)$ & $T$ (kg\,m$^2$\,s$^{-2}$) 
           & $z_T/(2a)$ & $\nu_d$ & $\nu_0$ \\
\hline
\hline
  I & 1562 & 100  & $3.07 \times 10^{-8}$ & 29.4 & 0.418 & 0.240 \\
\hline
 II & 3000 & 100  & $2.22 \times 10^{-8}$ & 21.2 & 1.110 & 0.396 \\
\hline
III & 1000 & 100  & $2.61 \times 10^{-9}$ & 2.49 & 3.151 & 0.567 \\
\hline
 IV & 1000 &  10  & $6.13 \times 10^{-9}$ & 5.85 & 13.41 & 0.755 \\
\hline
\end{tabular}
\end{center}
\caption{Simulation parameters for density profile measurements. In these
simulations, the particle radius $a=5 \times 10^{-4}$\,m and the particle 
mass $m=1.047 \times 10^{-6}$\,kg were not changed.}
\label{tab:simgpara}
\end{table}

The only way to get the correct theoretical density profile is for
points with $\nu < 0.65$. The value of $z_0=13.5 z_T$, where
$\nu(z_0/z_T)=0.65$, is taken from the simulation data and the
normalization accounts only for the $N_0$ particles above $z_0$.
The resulting density profile (dashed line) nicely agrees with
the simulations, and its limit of validity is inidcated by the angle
at $z/z_T=13.5$ and $\nu = 0.65$.

The reason for the increased density at low $z/z_T$ in the case of
simulation IV is the pressure drop due to crystallization in the 
equation of state, see Fig.\ \ref{fig:p2}. A higher density is
necessary to sustain a given pressure when $\nu > 0.65$. The data
for the lowest $z/z_T \approx 0$ are slightly off due to the wall
induced ordering at $z/z_T=0$.

\begin{figure}[thb]
  \center{
     ~\psfig{file=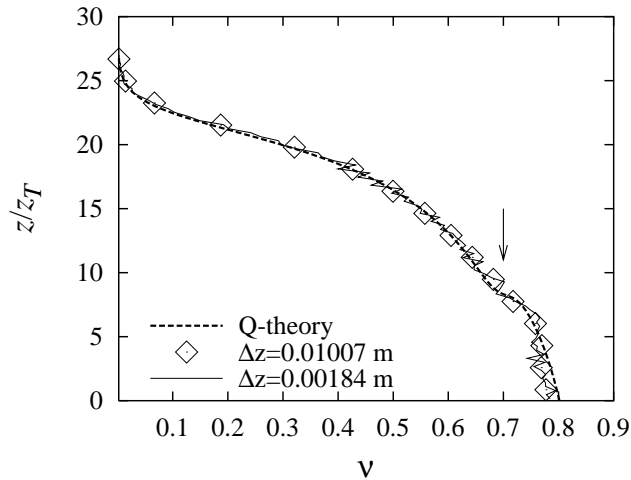,height=8.4cm,angle=-90,clip=}
     ~\vspace{-0.2cm}\\
    }
  \caption{Rescaled height $z/z_T$ plotted against the volume fraction
    $\nu$ from simulation IV, see table\ \protect\ref{tab:simgpara}.
    The symbols are simulation data of the particle-center with a
    rough binning $\Delta z=0.01007$\,m. The solid line is the density
    from the same data but with a much finer binning $\Delta z=0.00184$\,m.
    Both binnings start at $z_0+\frac{1}{5} a$ with $z_0=0$.
    The dashed line corresponds to the numerical solution of Eq.\
    (\protect\ref{eq:fgrav}) with $p=nT(1+Q_0)$. The transition density
    $\nu_c \approx 0.7$ is indicated by an arrow.
    }
  \label{fig:numsol}
\end{figure}

If, instead of $2\nu g_{2a}(\nu)$ in Eq.\ (\ref{eq:fgrav}), we use the 
more general form $Q_0$, we have to integrate
the differential equation $dp/dz=\nu m g /(\pi a^2)$ numerically with 
$p=nT(1+Q_0)$ and the condition that Eq.\ (\ref{eq:nu0def}) is fulfilled.
Simulation IV and the numerical solution are compared in Fig.\ 
\ref{fig:numsol}.  The qualitative behavior of the density profile
is well reproduced by the numerical solution with $\nu(z_0=0)=0.8016$.
Note that the averaging result is dependent on the binning -- we evidence
strong coarse-graining effects in the dense, ordered region with densities
$\nu > 0.70$.
~\\
~\\
~\\

\begin{figure}[htb]
\begin{center}
\psfig{file=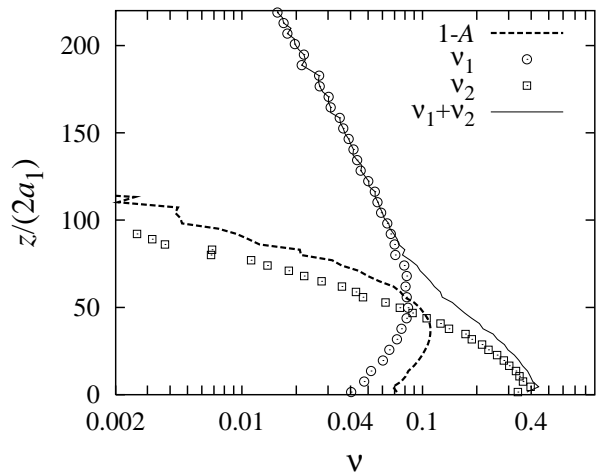,height=8.4cm,angle=-90}
\end{center}
\vspace{-0.9cm}~\\
\caption{Dimensionless height $z/(2a_1)$ plotted against the
volume fraction (in logarithmic scale). The thick, dashed line gives
$1-{\cal A}$ (which is zero when only the small species is present
at large heights), while the solid line indicates the total
volume fraction of the mixture $\nu^{\rm m}=\nu_1+\nu_2$.
}
\label{fig:bigrav}
\end{figure}

\subsection{Bidisperse systems with gravitation}

In Fig.\ \ref{fig:bigrav} the species volume fractions $\nu_1$ and
$\nu_2$ are plotted against the vertical coordinate $z/(2a_1)$. The data
are obtained after long equilibration, in a system with $N=2000$ particles,
width $L_x=l_x/(2a_1)=100$, and the size distribution of set D1 in the
previous section, with $a_1=0.0005$\,m.  The particles are no longer
mixed, as in the homogeneous, periodic systems of the previous section.
Segregation takes place, the larger and heavier particles (squares)
settle close to the bottom, whereas the gas of small and lighter particles
(circles) extends to larger heights.  Knowing both volume fractions, 
one could compute ${\cal A}$ as a function of the height, and insert it
into Eq.\ (\ref{eq:geff}) in order to compute the density profile.

\section{Summary and Outlook}
\label{sec:concl}

In summary, we reviewed existing theories for dilute and dense almost
elastic, smooth, 2D granular gases. For mono-, bi-, and polydisperse 
systems, we compared theoretical predictions with numerical simulations
of various systems. The collision frequency, the energy dissipation
and the equation of state, i.e.~the scaled pressure, are nicely
predicted by the theoretical expressions up to intermediate densities.
Especially, for arbitrary particle size distribution functions, the 
equation of state can be written in a nice form which only contains the 
width-correction ${\cal A}$ of the size-distribution function.
A small, empirical correction can be added to the theories to raise the 
quality even further. Finally, a merging function that connects the low 
and high density theories is proposed to give a {\em global equation of 
state} for {\em all} densities and size distribution functions. 

The equation of state is used to compute analytically and numerically 
the density profile of an elastic, monodisperse granular gas in the 
gravitational field. When a mixture is simulated, segregation is observed,
a case to which the theory cannot be applied. 
For densities below $\nu \approx 0.65$, the analytical solution 
works well, for higher densities close to the maximum density, one
has to use a numerical solver, since the global equation of state
cannot be integrated analytically. The strange shape of the density 
profile, as obtained from simulations, is nicely reproduced.

The simulations and the theories presented here were applied to homogeneous
systems. The range of applicability may be reduced by the fact that already
weak dissipation can lead to strong inhomogeneities in density, temperature, 
and pressure. In a freely cooling system, for example, clustering leads to
all densities between $\nu \approx 0$ and $\nu \approx \nu_{\rm max}$.
The proposed {\em global equation of state} is a necessary tool to 
account for such strong inhomogeneities with very high densities, above
which the low-density theory fails.  For another approach to handle
the high density regions, see Ref.~\cite{HongHere}.

The proposed global equation of state is based on a limited amount
of data. It has to be checked, whether it still makes sense in the extreme
cases of narrow $w(a)$, where crystallization effects are rather strong, 
and for extremely broad, possibly algebraic $w(a)$, where ${\cal A}$ is
not defined.  What also remains to be done is
to find similar expressions not only for pressure and energy dissipation
rate but also for viscosity and heat-conductivity and to extend the
theory to three dimensions

\section*{Acknowledgements}

We acknowledge the support by the Deutsche Forschungsgemeinschaft 
(DFG) and helpful discussions with B. Arnarson, D. Hong, J. Jenkins, 
M. Louge, and A. Santos.



\end{document}